\documentclass[epsfig,12pt]{article}
\pdfoutput=1
\usepackage{graphicx}
\usepackage{amsmath}
\usepackage{makeidx}
\usepackage{young}

\usepackage{a4,color,graphics}
\usepackage{amsfonts}
\usepackage{amssymb}
\usepackage{graphicx}%
\usepackage{epsfig}
\usepackage[all]{xy}
\input epsf.sty

\makeatletter \@addtoreset{equation}{section}

\def\be{\begin{equation}}
\def\ee{\end{equation}}
\def\bea{\begin{eqnarray}}
\def\eea{\end{eqnarray}}

\newcommand{\nc}{\newcommand}
\nc{\al}{\alpha} \nc{\bib}{\bibitem} \nc{\la}{\lambda}
\nc{\C}{\mbox{\hspace{1.24mm}\rule{0.2mm}{2.5mm}\hspace{-2.7mm}
C}} \nc{\R}{\mbox{\hspace{.04mm}\rule{0.2mm}{2.8mm}\hspace{-1.5mm}
R}}

\begin{document}\title{%
\rightline{\mbox {\normalsize {WITS-CTP-119 }}\bigskip}\textbf{  Orthogonal Schurs \\for Classical Gauge Groups   }}
\author{Pablo  Diaz\thanks{pablo.diazbenito@wits.ac.za}\\
{\small National Institute for Theoretical Physics, University
of Witwatersrand, South Africa}} \maketitle
\thispagestyle{empty}
\begin{abstract}
\bigskip
Finite $N$ physics of half-BPS operators for gauge groups $SO(N)$ and $Sp(N)$ has recently been studied\cite{CDD,CDD2}. Among other things they showed that, alike $U(N)$,  Schur operators (but in the square of their eigenvalues) diagonalize the free field two-point function of half-BPS operators for $SO(N)$ and $Sp(N)$ gauge groups. This result was unexpected since Wick contractions behave differently. In this paper we solve the puzzle by treating all gauge groups in a unified framework and showing how orthogonality of Schur operators emerges naturally from the embedding structure of classical Lie algebras $\mathfrak{g}(N)\hookrightarrow \mathfrak{g}(M)$. We go further and we state that orthogonality of Schurs is a gauge group-independent property for classical gauge groups.   \\\textbf{Keywords}: Schur polynomials, Weingarten functions, orthogonality, half-BPS operators, conformal field theories.
\end{abstract}
\newpage
\tableofcontents
\newpage
\pagestyle{plain}

\section{Introduction}

Recently, a finite $N$ physics for half-BPS operators in $\mathcal{N}=4$ super Yang-Mills with $SO(N)$ and $Sp(N)$ gauge groups have been studied\cite{CDD,CDD2} in the same spirit as finite physics for $U(N)$ was studied in \cite{CJR}. In \cite{CDD2} the authors realized that Schur operators (in the square of the eigenvalues) also diagonalized the free field two-point function for orthogonal and symplectic gauge groups. This was surprising. It is not obvious that Schur operators should form an orthogonal basis for $SO(N)$ and $Sp(N)$ gauge groups, since Wick contractions for these groups are completely different from the unitary case. \\
In this work we solve this puzzle by showing how Schur operators are naturally singled out as an orthogonal basis for classical gauge groups by the embedding structure of the Lie algebras. This way, the orthogonality of Schur operators should not be thought as associated to a specific gauge group but to a more essential embedding structure that classical algebras share. In other words, orthogonality of Schurs is a gauge group-independent property. These statements rely on the two main results of this paper: first, the compatibility of Weingarten calculus \cite{W,C,CM, Mat} with Wick calculus, explicitly expressed in equation (\ref{compatibility}); second, a `hidden' property of Schur operators, which turn out to be a  non-degenerate set of eigenvectors of self-adjoint operators naturally constructed from the embedding structure of the algebras (equation (\ref{eigenvectors})). Actually, this last property is intrinsic to Schur polynomials, it can be found  right from the context of symmetric functions\cite{OO1,OO2}.\\ 

According to AdS/CFT duality\cite{Mal,GKP,Wi}, we can learn string physics from studying the dual gauge theory, we just need the appropriate dictionary. In general, for a gauge group $G(N)$ in the the gauge theory, the limit $N\to \infty$ corresponds to supergravity solutions on the string side. For finite $N$, probing physics of the gauge theory\cite{BBNS} corresponds to studying non-perturbative objects such as Giant Gravitons\cite{MST,GMT,HHI}, as well as aspects of spacetime geometry captured in the stringy exclusion principle\cite{MS}.\\
It is known that different gauge groups in $\mathcal{N}=4$ super Yang-Mills correspond to different geometries in which the string theory lives. For $U(N)$ it is known that the corresponding background is $\text{AdS}_5\times\text{S}^5$ ~\cite{Mal}, while for $SO(N)$ and $Sp(N)$ gauge groups the CFT is dual to $\text{AdS}_5\times \mathcal{R}\text{P}_5$ geometry\cite{Wi2}. Unlike string physics in $\text{AdS}_5\times\text{S}^5$, strings in the orbifold $\text{AdS}_5\times \mathcal{R}\text{P}_5$ are non-oriented. From this fact one expects the study of non-perturbative stringy physics in the orbifold to bring new insights\cite{AABF}, and they are captured by the finite $N$ physics of the CFT with gauge groups $SO(N)$ and $Sp(N)$. This was one of the motivations for the study of finite $N$ physics of orthogonal and symplectic gauge groups in \cite{CDD,CDD2}.\\
The programme of studying finite $N$ physics in the case of unitary groups was initiated in \cite{CJR} for half-BPS operators, that is, for operators built on a single complex matrix. They showed that half-BPS operators can be described by Schur operators and they demonstrated that Schur operators diagonalize the free field two-point function. There has been a considerable progress on the study of finite $N$ physics for $U(N)$ gauge groups and, by now, we know a number of bases that diagonalize  the free field two-point function\cite{DSS,KR,BHR1,BHR2,BCD,K1,K2}. Orthogonal operators for the $Sl(2)$ sector of the theory, which involves gauge fields and their derivatives, and the action of the dilatation operator on them has been studied in \cite{DDS}. Fermion together with boson fields have been treated in \cite{DDN}, where an orthogonal restricted Schur basis for the whole $\frac{1}{8}$-BPS sector was found and the action of the dilatation operator on them described. We also know how to diagonalize the one-loop dilatation operator\cite{DMP,CDJ} for certain (large) operators dual to Giant Gravitons\cite{BBFH,CDL,DKS,DDGM, DR}. The diagonalization of the one-loop dilatation operator has provided new integrable sectors in the non-planar regime, with the spectrum of the dilatation operator reduced to that of decoupled harmonic oscillators which describe the excitations of strings attached to Giants\cite{DKS,DDGM,DR}.   \\
The  programme of studying finite $N$ physics in the case of orthogonal and symplectic groups was initiated in \cite{AABF}. A detailed study of the planar spectral
problem of  $\mathcal{N}=4$ super Yang-Mills with gauge groups SO(N) and Sp(N) was carried
out in \cite{CKZ}. In order to tackle the non-planar regime in future works, exact correlators of an orthogonal basis of half-BPS operators have been recently found in \cite{CDD,CDD2}. Among other things, \cite{CDD2} showed that for $SO(N)$ and $Sp(N)$ gauge groups Schur operators (although in the square of their eigenvalues) form a basis of  half-BPS operators and diagonalize the free field two-point function. As in the case of $U(N)$, their correlators were found to be given by nice (but different) combinatorial functions. In this paper we will call them $f_R^{G(N)}$, using a unified notation, where $G$ stands for the gauge groups $U$, $SO$ or $Sp$.\\

 As said above, it  was unexpected to find an orthogonal basis of Schurs for $SO(N)$ and $Sp(N)$ gauge groups since Wick contractions behave differently from the $U(N)$ case.  This paper solves the question of why Schur operators diagonalize the free field two-point function for classical gauge groups\footnote{The author would like to thank Robert de Mello Koch for pointing out the relevance of this question.}.
We show how the underlying structure of embedding algebras, common to all classical Lie algebras, naturally singles out Schurs as an orthogonal basis of half-BPS gauge invariant operators. Half-BPS gauge invariant operators for any gauge group $G(N)$ are built on an element (a complex matrix) $Z\in \mathfrak{g}(N)$.   We want to emphasize that orthogonality of Schur operators is not linked to the particular gauge group we are considering but to the embedding properties of the algebras, that is, the way $\mathfrak{g}(N)\hookrightarrow \mathfrak{g}(M)$.\\
The starting point is the infinite chain of embedding algebras
\begin{equation*}
\mathfrak{g}(1)\hookrightarrow \mathfrak{g}(2)\hookrightarrow \cdots
 \end{equation*}
This embedding has to be understood as follows. For $M>N$, an element $Z\in \mathfrak{g}(N)$ can be embedded in $\mathfrak{g}(M)$ by placing $Z$ in the upper left part of a matrix and fill the rest of the matrix, till dimension $M$, with 0's. One can easily check that the $M$-dimensional matrix so constructed is an element of $\mathfrak{g}(M)$, for $\mathfrak{g}$ a classical Lie algebra.\\
The next step is to design the most basic nontrivial set of operators that adapt to this structure. With this purpose we may define $\text{Proj}_{NM}$ as operators which act on any $Z\in\mathfrak{g}(M)$ by killing the `extra' 0's and lowering the dimension of the matrix to $N$. In the case that   $Z\in\mathfrak{g}(M)$ has more than $N$ eigenvalues different from 0, $\text{Proj}_{NM}$ takes it to $0_N$ matrix.\\
We may easily extend the definition of $\text{Proj}_{NM}$ to act on gauge invariant operators. We will make  $\text{Proj}_{NM}$  act on every $Z$ of the composite operators as above. This way, $\text{Proj}_{NM}$ is a map that takes gauge invariant operators built on $Z\in \mathfrak{g}(M)$ to gauge invariant operators built on $Z\in \mathfrak{g}(N)$.\\
At this stage and by virtue of the operator-state one-to-one correspondence in conformal field theories, it is useful to think of gauge invariant operators built on $Z\in \mathfrak{g}(N)$ as vectors belonging to a vector space $V_N$. Then $\text{Proj}_{NM}:V_M\to V_N$,  and the free field two-point function is an inner product defined in each $V_N$.  Now, each operator $\text{Proj}_{NM}$ must have an adjoint operator with respect to the inner products. We will call $\text{Av}_{MN}$ the adjoint operator of   $\text{Proj}_{NM}$. They are constructed in section \ref{main} by integrating over the gauge group the adjoint action of the gauge group onto the algebra.\\
Once we have a set of operators  and their adjoints we can construct a set of self-adjoint operators by composition. Operators  $\text{Proj}_{NM}\circ\text{Av}_{MN}$ are maps between gauge invariant operators built on $Z\in \mathfrak{g}(N)$. It turns out that the eigenvectors $\text{Proj}_{NM}\circ\text{Av}_{MN}$  are Schurs and their eigenvalues are all different. It follows immediately that Schurs are an orthogonal set of gauge invariant operators built on $Z\in \mathfrak{g}(N)$, for all $N$, with respect to the two-point function. \\

The paper is organized as follows. Section \ref{mathpre} aims to set the notation and to collect the mathematical tools necessary to follow the calculations. The notation we use in this work is slightly different from the notation used in previous papers. Basically, since most of the expressions relate objects with different rank of the gauge group we must specify it. So, changes in notation are mainly additions of a label for the rank parameter. The name `Schurs' for our operators and the notation we use for them in all classical groups is justified in subsection \ref{Schurs} by showing that the operators are actually Schur polynomials in the eigenvalues of $Z$ (for unitary groups) or in the square of its eigenvalues (for orthogonal and symplectic groups). \\
Our calculations for the unitary groups reduce at the end of the day to sums of characters of the symmetric group $S_n$ and their associated Schur polynomials. In the orthogonal and symplectic cases the relevant symmetries of the correlators are no longer those of class functions but of constant functions on the double coset of $S_{2n}$. They are generically called spherical functions and their associated symmetric polynomials go under the name of zonal polynomials. They play an analogous role for orthogonal and symplectic groups as characters and Schur polynomials do for the unitary cases. In subsection \ref{SF} we enter this analysis of symmetries to motivate the framework when we deal with each gauge group. We also collect some standard definitions and  results  (mainly from \cite{M}) which will be needed later on. \\
Combinatorial functions $f_R^G$ which were found in \cite{CJR} for unitary gauge groups and in \cite{CDD,CDD2} for orthogonal and symplectic gauge groups can be expressed as specific symmetric polynomials when evaluated at 1 in all the variables. We will need these definitions in section \ref{main} to state and prove the main results of the paper. So, we devote subsection \ref{fRfunctions} for this purpose.\\
The computation of the averaging operator acting on a gauge invariant operator involves integrals of matrix entries over gauge groups. These integrals have been computed for classical gauge groups and for finite $N$ over the last decade. The method to compute them was named Weingarten calculus after the paper \cite{C} and the functions involved Weingarten functions. We collect them in subsection \ref{Wgfunctions}.  \\
Section \ref{main} contains the main results of the paper, which are equations (\ref{compatibility}) and (\ref{eigenvectors}). In subsection \ref{GM} we flesh out the general method we have sketched in this introduction and discuss till which extend we can recover the two-point function of Schurs from (\ref{compatibility}) and (\ref{eigenvectors}). In the subsequent subsections we just prove (\ref{compatibility}) and (\ref{eigenvectors}) for each gauge group.\\
Finally, we summarize the results of the paper and outline some ideas for future works  in section \ref{Con}.

\section{Notation and mathematical preliminaries}\label{mathpre}
Although the mathematical context for operators and correlators of unitary groups is known we find it convenient to review it together with the other groups in order to point out their analogies. This helps in the purpose of finding a broader and unifying framework.
\subsection{Notation}
The notation we will be using along this paper suffers little modifications and some additions as compared to previous
physics papers on the matter. The main changes are explained in this subsection.\\
For all the groups and through all this paper we will reserve $n$ to denote the number of fields in a given operator, and $N$ or $M$ for the ranks of the gauge groups. Because we are interested in comparing correlators with different gauge group ranks we are forced to make ranks explicit. In the case of traces, we will change
\begin{eqnarray*}
\text{Tr}(\sigma)=\delta^{i_1}_{i_{\sigma(1)}}\cdots\delta^{i_n}_{i_{\sigma(n)}}&\to& \text{Tr}_N(\sigma) \nonumber \\
\text{Tr}(\sigma Z)=Z^{i_1}_{i_{\sigma(1)}}\cdots Z^{i_n}_{i_{\sigma(n)}}&\to& \text{Tr}_N(\sigma Z)\nonumber 
\end{eqnarray*}
to indicate that the indices which are traced run from $1,\dots,N$.
For operators we will also need to specify the rank of the group. So, for Schur operators we will change
\begin{equation*}
\chi_R(Z)=\frac{1}{n!}\sum_{\sigma\in S_n}\chi_R(\sigma)\text{Tr}(\sigma Z)\to \chi_{R,N}(Z)=\frac{1}{n!}\sum_{\sigma\in S_n}\chi_R(\sigma)\text{Tr}_N(\sigma Z).
\end{equation*} 
Besides, since the orthogonal basis of operators found in \cite{CDD} and \cite{CDD2} are also Schur polynomials (in the square of the eigenvalues) for the orthogonal and symplectic groups, we will refer to them as
\begin{equation*}
 \chi_{R,N}(Z^2)=\frac{1}{(n/2)!}\sum_{\sigma \in S_{n/2}}2^{-l(\sigma)}\chi_{R}(\sigma)\text{Tr}_N\big{(}\sigma (Z^2)^{\otimes \frac{n}{2}}\big{)},\quad R\vdash n/2.
\end{equation*}
Correlators of Schurs are given by the combinatorial  functions $f_R$, see (\ref{fRs}). For them, we will also need to specify the group, so we will write
\begin{equation*}
f_R^{U(N)}, \quad f_R^{SO(N)}\quad \text{and}\quad f_R^{Sp(N)}.
\end{equation*}
The usual short hand notation for tensors will be profusely used along the calculations. We will call
\begin{equation*}
Z^I_J\equiv Z^{i_1}_{ j_1}\cdots Z^{i_n}_{ j_n}, \quad
\delta^I_J\equiv \delta^{i_1}_{ j_1}\cdots \delta^{i_n}_{ j_n}, \quad
(\sigma)^I_J\equiv \delta^{i_1}_{ j_{\sigma(1)}}\cdots \delta^{i_n}_{ j_{\sigma(n)}}
\end{equation*}
and
\begin{equation*}
\text{Tr}_N(\sigma)\equiv  (\sigma)^I_I, \quad i_r=1,\dots,N.
\end{equation*}
For the orthogonal and symplectic cases we will use the short hand notation
\begin{equation*}
\delta_I\equiv\delta_{i_1 i_2}\cdots \delta_{i_{2n-1} i_{2n}},\quad Z_I\equiv Z_{i_1 i_2}\cdots Z_{i_{2n-1} i_{2n}},
\end{equation*}
and also
\begin{equation*}
\delta_{\alpha(I)}\equiv \delta_{i_{\alpha(1)} i_{\alpha(2)}}\cdots \delta_{i_{\alpha(2n-1)} i_{\alpha(2n)}}.
\end{equation*}
For symmetric functions it will often be the case that we need their value when all the variables are equal to one. In this case we will write
\begin{equation*}
p_\sigma({\bf 1}_N)\equiv p_\sigma(\underbrace{1,1,\dots,1}_N).
\end{equation*}

\subsection{Why `Schurs'?}\label{Schurs}
The name of `Schur operators' for the orthogonal basis of operators found in \cite{CJR} in the case of unitary groups, and in \cite{CDD2} for orthogonal and symplectic groups, is justified because the operators are actually Schur polynomials whose variables are the eigenvalues of the element of the algebra $Z\in \mathfrak{g}$ that builds the operator.\\
 Let us first see how it works for unitary groups.
Gauge invariant operators built from $Z$'s are easily seen to be symmetric functions in the eigenvalues of $Z$.
Since $\text{Tr}$ is a basis-invariant operation, we can choose a basis of eigenvectors of $Z$ and see that
\begin{equation*}
 \text{Tr}_N(\sigma Z)=p_{\sigma}(\lambda_1,\dots,\lambda_N).
\end{equation*}
Although Schur polynomials can be constructed in different ways, we find convenient to write them in association with characters by the relation
\begin{equation}\label{defSchur}
s_R(x_1,\dots,x_N)=\frac{1}{n!}\sum_{\sigma \in S_n}\chi_R(\sigma)p_{\sigma}(x_1,\dots,x_N).
\end{equation} 
Thus
\begin{equation*}
\chi_{R,N}(Z)\equiv\frac{1}{n!}\sum_{\sigma \in S_n}\chi_R(\sigma)\text{Tr}_N(\sigma Z)=s_R(\lambda_1,\dots,\lambda_N),\quad R\vdash n,
\end{equation*}
where $\{\lambda_1,\dots,\lambda_N\}$ are the eigenvalues of $Z$.\\
For the orthogonal and symplectic algebras, an orthogonal basis of operators was found in \cite{CDD2} to be
\begin{equation}\label{defnormal}
\chi_{R,N}(Z^2)=\frac{1}{(n/2)!}\sum_{\sigma \in S_{n/2}}2^{-l(\sigma)}\chi_{R}(\sigma)\text{Tr}_N\big{(}\sigma (Z^2)^{\otimes \frac{n}{2}}\big{)},\quad R\vdash n/2.
\end{equation}
Using the invariance of $\text{Tr}$ under change of basis we can write
\begin{equation}\label{Zeigenvalues}
Z=\left(
\begin{array}{ccccc}
 & \lambda_1 &  &&\\
-\lambda_1& &&&  \\
&  & \ddots&& \\
 &  &  &&\lambda_{N/2}\\
 & &  &-\lambda_{N/2}&
\end{array} \right), 
\end{equation}
for $Z\in \mathfrak{so}(N)$. In the case of $N$ odd, for $Z\in \mathfrak{so}(N)$ we just have to add a 0 eigenvalue. 
For all these algebras, $Z^2$ is diagonal in the square of their eigenvalues. It is easy to see that for $N$ even a multitrace monomial can be expressed in terms of symmetric power sums as
\begin{equation*}
\text{Tr}_N\big{(}\sigma (Z^2)^{\otimes \frac{n}{2}}\big{)}=(-1)^{\frac{n}{2}}2^{l(\sigma)}p_{\sigma}(\lambda_1^2,\dots, \lambda_{N/2}^2),
\end{equation*}
and if $Z\in \mathfrak{so}(N)$ with $N$ odd
\begin{equation*}
\text{Tr}_N\big{(}\sigma (Z^2)^{\otimes \frac{n}{2}}\big{)}=(-1)^{\frac{n}{2}}2^{l(\sigma)}p_{\sigma}(\lambda_1^2,\dots, \lambda_{\frac{N-1}{2}}^2,0)=(-1)^{\frac{n}{2}}2^{l(\sigma)}p_{\sigma}(\lambda_1^2,\dots, \lambda_{\frac{N-1}{2}}^2).
\end{equation*}
So, our operators read
\begin{eqnarray*}
\chi_{R,N}(Z^2)&=&\frac{1}{(n/2)!}\sum_{\sigma \in S_{n/2}}2^{-l(\sigma)}\chi_{R}(\sigma)\text{Tr}_N\big{(}\sigma (Z^2)^{\otimes \frac{n}{2}}\big{)}\nonumber \\
&=&(-1)^{\frac{n}{2}}\frac{1}{(n/2)!}\sum_{\sigma \in S_{n/2}}2^{-l(\sigma)}\chi_{R}(\sigma)2^{l(\sigma)}p_{\sigma}(\lambda_1^2,\dots, \lambda_{N/2}^2)\nonumber \\
&=&(-1)^{\frac{n}{2}}s_R(\lambda_1^2,\dots, \lambda_{N/2}^2), \quad R\vdash n/2.
\end{eqnarray*}
This means that, up to an irrelevant global minus factor, the operators $\chi_{R,N}(Z^2)$ are actually Schur polynomials in their eigenvalues. This justifies the name and the notation.

\subsection{Symmetries and framework}\label{SF}
The mathematics which appear in our calculations involve broad and quite developed fields.
A complete introduction to them is out of the scope of this paper. Instead, we will
offer some intuitive justification of the functions and polynomials involved for each 
gauge group and the necessary formulae and definitions to follow the calculations. We will
give some references where the interested reader can find a full description.\\
\paragraph{Unitary groups.}
When we work on the unitary case, say $Z\in \mathfrak{u}(N)$, a general monomial multitrace operator with $n$ fields
\begin{equation*}
\text{Tr}_N(\sigma Z)=Z^I_{\sigma(I)}, \quad \sigma\in S_n
\end{equation*}
depends only on the cycle structure of $\sigma$.  So, every element in the conjugate class of $\sigma$ will lead to the same 
operator. For this reason, the functions that appear in the unitary context are class functions, for which the characters
are a standard basis. For example, correlators
\begin{equation*}
\langle \text{Tr}_N(\sigma Z)\text{Tr}_N(\tau \bar{Z})\rangle
\end{equation*}
are manifestly class functions on the elements $\sigma, \tau \in S_n$. Therefore, correlators can be expanded
in terms of characters of $S_n$, in fact
\begin{equation} \label{expandU}
\langle \text{Tr}_N(\sigma Z)\text{Tr}_N(\tau \bar{Z})\rangle=\sum_{\rho\in S_n}\text{Tr}_N(\tau\rho\sigma\rho^{-1})=\sum_{\lambda\vdash n} f^{U(N)}_\lambda \chi_\lambda(\tau)\chi_\lambda(\sigma).
\end{equation}
Note that, except for the value of $f^{U(N)}_\lambda$, formula  (\ref{expandU}) could be almost guessed by symmetry considerations.\\
Other class functions that often appear in the unitary case are traces, and they also can be expanded in terms of characters as
\begin{equation}\label{traceU}
\text{Tr}_N(\sigma)=\frac{1}{n!}\sum_{\lambda\vdash n}d_\lambda f^{U(N)}_\lambda\chi_\lambda(\sigma).
\end{equation} 
We can associate symmetric polynomials to functions on the symmetric group. Characters are associated with Schur polynomials by relation (\ref{defSchur}).\\
Later, we will need the value of Schur polynomials when all the variables are equal to 1 since they are related to the combinatorial functions $f_R^{U(N)}$ , see (\ref{fRU}). This specialization of Schur polynomials is
\begin{equation*}
s_R({\bf 1}_N)=\frac{d_R}{n!}\prod_{(i,j)\in R}(N+j-i).
\end{equation*}
 

\paragraph{Orthogonal groups.}
Elements $Z\in \mathfrak{so}(N)$ must fulfills the relation\footnote{Note that, unlike the unitary group, the transposition of any of the matrices 
$Z\in \mathfrak{so}(N)$  in a given multitrace operator still gives  a holomorphic operator. So, it is a relevant symmetry for us.} $Z^t=-Z$.  
This forces any trace within a multitrace monomial to have an even number of fields. Clearly the monomials
\begin{equation} \label{monomialZ2}    
\text{Tr}_N\big{(}\sigma (Z^2)^{\otimes \frac{n}{2}}\big{)}\qquad R\vdash n/2
\end{equation}
generate all half-BPS multitrace gauge invariant operators. Monomials (\ref{monomialZ2}) resemble the shape of  multitrace monomials for unitary groups as we change $Z\to Z^2$ and $n\to n/2$. One
might be tempted to establish some relations between orthogonal and unitary groups upon this fact. For example, one can argue that correlators
\begin{equation}\label{corrO}
\langle\text{Tr}_N\big{(}\sigma (Z^2)^{\otimes \frac{n}{2}}\big{)}\text{Tr}_N\big{(}\tau (\bar{Z}^2)^{\otimes \frac{n}{2}}\big{)}\rangle
\end{equation}
are class functions of $\sigma$ and $\tau$ and may be expanded in terms of characters of $S_{n/2}$. This is indeed the case.
 However, as soon as we try to flesh it out and compute any correlator it turns out that Wick contractions act on elements of the algebra, and $Z^2$ is neither an element of $\mathfrak{so}(N)$ nor an element of  $\mathfrak{u}(N)$. The same happens when we try to implement Weingarten calculus, as we are doing later on. In the latter case, it happens that the adjoint action of the group is defined on the algebra, and again, $Z^2$ is not an element of the algebra.\\
 Then, it is convenient to write (\ref{monomialZ2}) in a different way. \\
For this purpose let us call $w=(1234)(5678)\cdots(2n-3~2n-2~ 2n-1~ 2n)$. And let $\sigma$ still be a permutation of $n/2$ elements but on the set of numbers $\{1,5,\dots,2n-3\}$.
 We will also use the notation $w_{\sigma} \equiv\sigma w$, so $w_{\sigma}$ is an element of $S_{2n}$ built from $w$ by joining its cycles according to $\sigma$. One can see that\footnote{There is an abuse of notation in equation (\ref{monomialZ}). On the LHS $\sigma$ permutes elements of the set $\{1,2,\dots, n/2\}$ while on the RHS it does for elements of $\{1,5,\dots,2n-3\}$. Being aware of this, we keep the notation for simplicity since both sides of the equation depend only on the cycle structure of $\sigma$.} 
\begin{equation}\label{monomialZ}
\text{Tr}_N\big{(}\sigma (Z^2)^{\otimes \frac{n}{2}}\big{)}=\delta_I Z_{w_{\sigma}(I)}.
\end{equation} 
Now, on the RHS of (\ref{monomialZ}), $Z$'s appear `unpaired'. It is immediate to realize that although the monomials are still invariant under conjugation of $\sigma$ they are not invariant under conjugation of of $w_{\sigma}\in S_{2n}$. With respect to $w_{\sigma}$ they develop a different symmetry, as we are going to see.\\
 The hyperoctahedral group $S_n[S_2]$ is a subgroup of $S_{2n}$. If we fix  couples in the set of number on which $S_{2n}$ acts, say $\{1,2; 3,4;\cdots ;2n-1,2n\}$, then $S_n[S_2]$  consists of permutations of $S_{2n}$ that move couples and can, additionally, flip the order of each couple. It is easy to see that $|S_n[S_2]|=|S_2^{\otimes n}||S_n|=2^nn!$.  \\
From (\ref{monomialZ}) we see that $\text{Tr}_N\big{(}\sigma (Z^2)^{\otimes \frac{n}{2}}\big{)}$ is (up to a sign) constant on the elements of the double coset of $w_{\sigma}$ , that is
\begin{equation}\label{symmetryO}
\delta_I Z_{\xi w_{\sigma}\eta(I)}=
\delta_I Z_{w_{\sigma}(I)}\text{sgn}(\eta),\quad \xi,\eta \in S_n[S_2].
\end{equation}
Consequently, correlators like (\ref{corrO}) may be expressed in terms of constant (up to a sign) functions on the double coset of 
$w_{\sigma}$. Constant functions on the double coset $S_n[S_2]\setminus S_{2n}/ S_n[S_2]$ naturally appear  in the context of Gelfand pairs $(S_{2n},S_n[S_2])$, see Chapter VII of \cite{M} for details.\\
We will make difference among three type of functions depending on how they behave on the elements of the double coset. Standard bases for them are 
\begin{eqnarray}\label{sphericalfunctions}
\omega_\lambda(\xi\sigma\eta)&=&\omega_\lambda(\sigma), \quad \lambda\vdash n\quad \text{(spherical functions)},\nonumber \\
\theta_{\lambda}(\xi\sigma\eta)&=&\theta_\lambda(\sigma)\text{sgn}(\eta), \quad \lambda\vdash n/2\quad \text{(bispherical functions)},\nonumber \\
\omega_{\lambda}^{\varepsilon}(\xi\sigma\eta)&=&\omega_\lambda^{\varepsilon}(\sigma)\text{sgn}(\xi\eta), \quad \lambda\vdash n \quad \text{(twisted spherical functions)}
\end{eqnarray}
where $\sigma \in S_{2n}$ and $\xi,\eta \in S_n[S_2]$. For the orthogonal case only functions $\theta$ and $\omega$ will be relevant.\\
Functions (\ref{sphericalfunctions}) are defined in terms of characters. For spherical  functions we have
\begin{equation}\label{omega}
\omega_{\lambda}(\sigma)=\frac{1}{2^n n!}\sum_{\xi \in S_n[S_2]}\chi_{2\lambda}(\sigma \xi), \quad \sigma \in S_{2n},\quad \lambda\vdash n.
\end{equation}
Spherical functions enjoy the orthogonality relations inherited from characters. \\
Associated with each $\omega_{\lambda}$ there is a symmetric polynomial $Z_{\lambda}$ (zonal polynomial).  Zonal polynomials are the analogues of Schur polynomials. In the same manner of Schur polynomials in (\ref{defSchur}), zonal polynomials are defined in terms of  spherical functions as  
\begin{equation}\label{zonal}
Z_{\lambda}(x_1,\dots, x_N)=\frac{ 2^n n!}{2n!}\sum_{\rho \in S_n}2^{-l(\rho)}\omega_{\lambda}(h_{\rho})p_{\rho}(x_1,\dots, x_N)
\end{equation}
where $S_{2n}\ni h_\rho=\rho (12)(34)\cdots(2n-1~2n)$, and $\rho\in S_n$ acting on the set of numbers $\{1,3,\dots,2n-1\}$. 
 Zonal polynomials have also a `cut off', that is
$Z_{\lambda}({\bf 1}_M)=0$ if $l(\lambda)>M$.\\
 The specializations of zonal polynomials to 1 in their variables are 
\begin{equation}\label{zonalspecialO}
Z_{\lambda}({\bf 1}_N)=\prod_{(i,j)\in \lambda}(N+2j-i-1).
\end{equation} 
Alike the $U(N)$ case, there is a direct relation between specializations of zonal polynomials and combinatorial functions $f_R^{SO(N)}$, see equation (\ref{fRO}).

Let us now talk about functions $\theta$. They where fully studied in \cite{I}. Before writing an explicit definition in terms of
characters we will see that the essence of the result found in \cite{I} can be understood in terms of pure symmetry considerations.\\
In view of  (\ref{monomialZ}) we can write the identity
\begin{equation}\label{id}
\langle\text{Tr}_N\big{(}\sigma (Z^2)^{\otimes \frac{n}{2}}\big{)}\text{Tr}_N\big{(}\tau (\bar{Z}^2)^{\otimes \frac{n}{2}}\big{)}\rangle=\langle\delta_I Z_{w_{\sigma}(I)}\delta_I \bar{Z}_{w_{\tau}(I)}\rangle.
\end{equation}
As argued before, the LHS of (\ref{id}) is manifestly a class function of $S_{n/2}$ and can be expanded in terms of characters $\chi_{\lambda}(\sigma)$ and $\chi_{\mu}(\tau)$, for $\lambda,\mu\vdash n/2$. The RHS of  (\ref{id}) is  constant (up to a sign) on the elements of the double coset of $w_\sigma$ and $w_\tau$, and can consequently be written in terms of $\theta_\lambda(w_\sigma)$ and $\theta_\mu(w_\tau)$, for $\lambda,\mu\vdash n/2$. Because $\sigma$ and $\tau$ are arbitrary there has to be a relation between both basis. That is, functions $\theta$ of elements of  $S_{2n}$ must be expressible in terms of characters of elements of $S_{n/2}$ and vice versa. To fix a basis $\{\theta_\lambda\}$ one must impose a condition. The condition which was implicitly imposed in \cite{I} is that the associated symmetric polynomials of $\theta$'s were Schur polynomials.   With this condition, functions $\theta$ were defined  as
\begin{equation*}
\theta_\lambda(w_\sigma)=\frac{2^{l(\sigma)}\sqrt{\text{Hooks}_{2(\lambda\cup\lambda)}}}{2^n n!}\chi_\lambda(\sigma),\quad \lambda\vdash n/2.
\end{equation*}
With a bit of work one can translate this result into an identity for characters
\begin{equation}\label{Ivanov}
\frac{d_{2(\lambda\cup \lambda)}}{2n!}\sum_{\xi, \eta \in S_n[S_2]}\chi_{2(\lambda\cup \lambda)}(\xi w_\sigma \eta w_\tau) \text{sgn}(\xi)=2^{l(\sigma)+l(\tau)}\chi_{\lambda}(\sigma)\chi_{\lambda}(\tau) \quad \lambda\vdash n/2,
\end{equation}
and 
\begin{equation}\label{Ivanov2}
\sum_{\xi, \eta \in S_n[S_2]}\chi_S(\xi w_\sigma \eta w_\tau) \text{sgn}(\xi)=0 \quad S\vdash 2n
\end{equation}
if $S$ has any row or column with an odd number of boxes. For later use, let us write (\ref{Ivanov}) and (\ref{Ivanov2})  in terms of spherical functions:
\begin{equation}\label{Ivanovw}
d_{2R}\frac{2^n n!}{(2n)!}\sum_{\xi \in S_n[S_2]}\omega_{R}(w_{\sigma}^{-1}\xi w_{\tau}^{-1})  \text{sgn}(\xi)=
\begin{cases}
&2^{l(\sigma)+l(\tau)}\chi_{\lambda}(\sigma)\chi_{\lambda}(\tau),~ \text{if} ~R=\lambda\cup\lambda\\
&0, ~\text{if}~ R~ \text{has odd number of boxes in any column}. 
\end{cases}
\end{equation}
This identity will be very useful later on.\\

A common trace that comes out often in the orthogonal case is
\begin{equation*}
\delta_{\alpha(I)}\delta^I=N^{\text{co}(\alpha)}, \quad \alpha \in S_{2n},
\end{equation*}
where $\text{co}(\alpha)$ is the coset-type of $\alpha$, which is a partition of $n$. This traces
are the analogue of $\text{Tr}(\alpha)$ in the unitary case. For the same reason that $\text{Tr}(\alpha)$ could be expanded
in terms of characters as in (\ref{traceU}), $\delta_{\alpha(I)}\delta^I$ may be expanded in terms of spherical functions.
It is known (see for example \cite{CM}) that 
\begin{equation}\label{traceO}
\delta_{\alpha(I)}\delta^I=\frac{2^n n!}{(2n)!}\sum_{\lambda \vdash n}d_{2\lambda}Z_{\lambda}({\bf 1}_N)\omega_{\lambda}(\alpha).
\end{equation}
Note that the sum in (\ref{traceO}) only involves partitions $l(\lambda)\leq N$ since zonal polynomials are otherwise 0.

\paragraph{Symplectic groups.}
The elements of the symplectic group $Sp(N)$ ($N$ even) must fulfill $g^{-1}=-Jg^tJ$ , where $J$ is the standard antisymmetric matrix 
\begin{equation}\label{J}
J=\left(
\begin{array}{c|c}
 0_{N/2}&I_{N/2}  \\
\hline
-I_{N/2}&0_{N/2}  
\end{array} \right). 
\end{equation}
Elements $Z$ of the algebra fulfill $JZ+Z^tJ=0$. 
\\
Because the operation $\text{Tr}(Z^k)$ is invariant under transposition  of $Z^k$, and $J^2=-1$, it is easy to see that the multitrace monomials must have, as in the orthogonal case, an even number of fields in each trace. So they can also be written as
\begin{equation*}
\text{Tr}_N\big{(}\sigma (Z^2)^{\otimes \frac{n}{2}}\big{)}=\delta_I Z_{w_{\sigma}(I)}.
\end{equation*} 
In order to see the symmetries of multitrace monomials we will rewrite
\begin{equation}\label{traceJ}
\delta_I Z_{w_{\sigma}(I)}=J_I(JZ)_{w_{\sigma}(I)}.
\end{equation}
Because $JZ$ is invariant under transposition, and $J$ are antisymmetric, we see that
\begin{equation}\label{symmetrySp}
J_I (JZ)_{\xi w_{\sigma}\eta(I)}=
J_I (JZ)_{w_{\sigma}(I)}\text{sgn}(\xi),\quad \xi,\eta \in S_n[S_2].
\end{equation}
Thus, multitrace monomials built on $Z\in\mathfrak{sp}(N)$ enjoy the same symmetry as for the orthogonal case.
Correlators of traces may be, again, expanded in terms of bispherical functions and also in terms of characters of $S_{n/2}$. In fact,
it was found in \cite{CDD2} that for $\sigma,\tau \in S_{n/2}$
\begin{equation}\label{coortraceSp}
\langle\text{Tr}_N(\sigma(Z^2)^{\otimes\frac{n}{2}})\text{Tr}_N(\tau(\bar{Z}^2)^{\otimes\frac{n}{2}})\rangle =\sum_{R\vdash n/2}2^{l(\sigma)+l(\tau)}\chi_R(\sigma)\chi_R(\tau)f_R^{Sp(N)},
\end{equation}
which is the same relation as for correlators of traces in the orthogonal group after the change $f_R^{Sp(N)}\leftrightarrow f_R^{SO(N)}$.\\
The relevant spherical functions involved in the symplectic case are $\theta$ and $\omega^{\varepsilon}$, which have been defined in (\ref{sphericalfunctions}). \\
Twisted spherical functions can also be expressed in terms of characters as
\begin{equation}\label{omegae}
\omega_{\lambda}^{\varepsilon}(\sigma)=\frac{1}{2^n n!}\sum_{\xi \in S_n[S_2]}\chi_{\lambda\cup\lambda}(\sigma \xi)\text{sgn}(\xi), \quad \sigma \in S_{2n},\quad \lambda\vdash n.
\end{equation}
Note that there is a simple relation between spherical and twisted spherical functions
\begin{equation*}
\omega_\lambda(\sigma)=\omega_{\lambda'}^\varepsilon(\sigma)\text{sgn}(\sigma).
\end{equation*}
Associated to twisted spherical functions are also symmetric polynomials: the so-called twisted zonal polynomials, defined as
\begin{equation}\label{zonalsp}
Z'_{\lambda}(x_1,\dots, x_N)= \frac{ 2^n n!}{2n!}\sum_{\rho \in S_n}2^{-l(\rho)}\omega^{\varepsilon}_{\lambda}(h_{\rho})p_{\rho}(x_1,\dots, x_N),
\end{equation}
whose specialization is
\begin{equation}\label{twistedzonalspecial}
Z'_{\lambda}({\bf 1}_N)=\prod_{(i,j)\in \lambda}(2N+j-2i+1).
\end{equation}
Again, the specialization of twisted zonal polynomials gives the value of the combinatorial functions $f_R^{Sp(N)}$, as explicitly written in equation (\ref{fRsPZ}).\\
A common trace that appears in the context of symplectic groups is
\begin{equation}\label{symplectic}
J_I J_{\alpha(I)}=\text{sgn}(\alpha)(-N)^{\text{co}(\alpha)},
\end{equation}
and this trace may also be expanded in terms of twisted spherical functions as
\begin{equation}\label{traceSp}
J_I J_{\alpha(I)}=\frac{2^n n!}{(2n)!}\sum_{\substack{\lambda \vdash n \\
l(\lambda)\leq N}}d_{\lambda\cup\lambda}Z'_{\lambda}({\bf 1}_{N/2})\omega^\varepsilon_{\lambda}(\alpha).
\end{equation}

\subsection{functions $f_R^G$}\label{fRfunctions}
Functions $f_R$ appear in the correlator of Schurs as 
\begin{eqnarray}\label{fRs}
\langle\chi_{R,N}(Z)\chi_{S,N}(\bar{Z})\rangle&=&f_R^{U(N)}\delta_{RS}\nonumber \\
\langle\chi_{R,N}(Z^2)\chi_{S,N}(\bar{Z}^2)\rangle&=&f_R^{SO(N)}\delta_{RS} \nonumber \\
\langle\chi_{R,N}(Z^2)\chi_{S,N}(\bar{Z}^2)\rangle&=&f_R^{Sp(N)}\delta_{RS}, \quad (N ~\text{even}).
\end{eqnarray}
For the unitary case  $f_R^{U(M)}$ was found in  \cite{CJR} to be
\begin{equation*}
f_R^{U(N)}=\frac{1}{d_R}\sum_{\sigma \in S_n}\chi_R(\sigma)\text{Tr}_N(\sigma)=\prod_{(i,j)\in R}(N+j-i).
\end{equation*}
But $\text{Tr}_N(\sigma)$ is not other thing that the specialization of the power sum with $N$ eigenvalues equal to 1, so,
in terms of symmetric functions
\begin{equation}\label{fRU}
 f_R^{U(N)}=\frac{1}{d_R}\sum_{\sigma \in S_n}\chi(\sigma)p_\sigma({\bf 1}_N)= \frac{n!}{d_R}s_R({\bf 1}_N).
\end{equation}
For the orthogonal case $f_R^{SO(N)}$  was given in \cite{CDD}.  We will give here the prescription, for a justification see \cite{CDD}.
If $R\vdash n/2$ is the label of the Schur operator\footnote{Note the difference in notation between ours an that of \cite{CDD,CDD2}. In  \cite{CDD,CDD2}, instead of $R\vdash n/2$, Schur operators for $SO(N)$ and $Sp(N)$ were labeled by $R\vdash 2n$ with the condition that $R=2\lambda \cup 2\lambda$. Both labelings refer to the same object. In this work we find our labels more convenient  since they lead to uniform expressions for all classical gauge groups and easier relations between functions $f_R$ and the specializations of symmetric polynomials in (\ref{fRO}) and (\ref{fRsPZ}), which will be needed in section \ref{main}.} as used in (\ref{fRs}) then one must blow each box of $R$ into 4-blocks. This is done by the operation $2R\cup 2R$, where $2R$ denotes the diagram constructed from $R$ by doubling the number of boxes in each row, and $R\cup R$ is obtained from $R$ by doubling the number of boxes in each column. Graphically
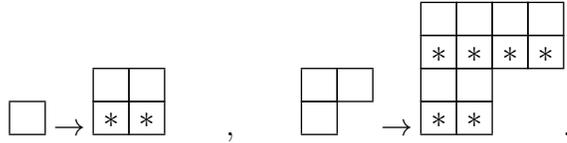
\begin{equation*}
\begin{Young}
\cr
\end{Young}
\to
\begin{Young}
&\cr
$*$&$*$\cr
\end{Young}\qquad , \qquad
\begin{Young}
&\cr
\cr
\end{Young}
\to
\begin{Young}
&&&\cr
$*$&$*$&$*$&$*$\cr
&\cr
$*$&$*$\cr
\end{Young}.
\end{equation*}
The stars mark the boxes we must consider in order to find
\begin{equation}\label{presO}
f_R^{SO(N)}=\prod_{(i ~\text{even},j)\in 2R\cup 2R}(N+j-i).
\end{equation}
But this could be written in a different form in order to make contact with symmetric functions. As before,
let us label our Schur operators with $R\vdash n/2$. 
It can be easily seen that 
\begin{equation}\label{fRO}
f_R^{SO(N)}=\prod_{(i,j)\in R\cup R}(N+2j-i-1)=Z_{R\cup R}({\bf 1}_N),
\end{equation}
in accordance with (\ref{zonalspecialO}).
\\
For the symplectic case an analogous prescription to (\ref{presO}) was given in \cite{CDD2}. As in the orthogonal case,
if $R\vdash n/2$ is the label of the Schur operators as used in (\ref{fRs}) then one must take $2R\cup 2R$ as
\begin{equation*}
\begin{Young}
\cr
\end{Young}
\to
\begin{Young}
$*$&$*$\cr
&\cr
\end{Young}\qquad , \qquad
\begin{Young}
&\cr
\cr
\end{Young}
\to
\begin{Young}
$*$&$*$&$*$&$*$\cr
&&&\cr
$*$&$*$\cr
&\cr
\end{Young},
\end{equation*}
but now, the boxes we have to consider are in odd rows. Thus
\begin{equation}\label{presSp}
f_R^{Sp(N)}=\prod_{(i ~\text{odd},j)\in 2R\cup 2R}(N+j-i).
\end{equation}
Again, it will be convenient to relate this formula to symmetric functions. It can be seen that
\begin{equation}\label{fRsPZ}
f_R^{Sp(N)}=\prod_{(i,j)\in 2R}(N+j-2i+1)=Z'_{2R}({\bf 1}_{N/2}),
\end{equation}
in accordance with (\ref{twistedzonalspecial}).
\subsection{Weingarten calculus}\label{Wgfunctions}
In section \ref{main} we will define the operators $\text{Av}_{MN}$ as integrals over the gauge group. The integrands are entries of the group matrices. It turns out that these integrals can be exactly computed  and lead to nice combinatorics involving symmetric functions. \\
Weingarten was the first in trying to compute them, and he succeeded for the asymptotic behaviour\cite{W}, that is, for large $N$.
Since 2003 on, these integrals have been computed for finite $N$ and for all the classical gauge groups, see \cite{C,CM,Mat}. The method for computing them has been baptized as `Weingarten calculus', and the combinatorial functions involved `Weingarten functions'. For its close relation with random matrix theory, Weingarten calculus has been widely applied in several fields of mathematics and physics, but as far as we know this is the first time it appears in our context.\\
Here we present the formulas of the Weingarten functions for the different gauge groups which will be necessary to follow the calculations of section \ref{main}.
\\
On the course of our computations in subsection \ref{caseU}, we will need the Weingarten functions associated with the unitary group. They are defined as\cite{C}
\begin{equation}\label{WgU}
\text{Wg}^{U(N)}(\sigma)=\frac{1}{n!^2}\sum_{\substack{\lambda \vdash n \\
l(\lambda)\leq N}}\frac{d_{\lambda}^2}{s_{\lambda}({\bf 1}_N)}\chi_{\lambda}(\sigma)=\frac{1}{n!}\sum_{\substack{\lambda \vdash n \\
l(\lambda)\leq N}}\frac{d_{\lambda}}{f_\lambda^{U(N)}}\chi_{\lambda}(\sigma),\quad \sigma\in S_n.
\end{equation}
The condition $l(\lambda)\leq N$ is necessary because Schur polynomials have a natural `cut off'. They are 0 if the of parts of $\lambda$ 
 exceeds the number of variables $N$. This would create a pole in (\ref{WgU}) and the Weingarten function would be ill-defined.
\\
In subsection \ref{caseO}, when dealing with orthogonal gauge groups we will need their Weingarten function.
 They are defined as\cite{CM}
\begin{equation}\label{WgO}
\text{Wg}^{O(N)}(\alpha)=\frac{2^n n!}{(2n)!}\sum_{\substack{\lambda \vdash n \\
l(\lambda)\leq N}}d_{2\lambda}\frac{\omega_{\lambda}(\alpha)}{Z_{\lambda}({\bf 1}_{N})}, \quad \alpha \in S_{2n}.
\end{equation}   
Note that the condition $l(\lambda)\leq N$ is also necessary because Zonal polynomials have a natural `cut off'. They are 0 if the of parts of $\lambda$  exceeds the number of variables $N$.
\\
The Weingarten function for the symplectic case can also be found in \cite{Mat} and it reads
\begin{equation}\label{WgSp}
\text{Wg}^{Sp(N)}(\alpha)=\frac{2^n n!}{(2n)!}\sum_{\substack{\lambda \vdash n \\
l(\lambda)\leq N}}\frac{d_{\lambda\cup\lambda}}{Z'_\lambda({\bf 1}_{N/2})}\omega^{\varepsilon}_\lambda (\alpha), \quad \alpha\in S_{2n},
\end{equation}
and, again,  $l(\lambda)\leq N$ needs to hold for the function to be well-defined.

\section{Orthogonality of Schurs for classical gauge groups}\label{main}
This section contains the main results of the paper which are succinctly expressed in equations (\ref{compatibility}) and (\ref{eigenvectors}).

\subsection{General methodology}\label{GM}
First, let us consider the infinite embedding chain
\begin{equation}\label{embebding}
\mathfrak{g}(1)\hookrightarrow \mathfrak{g}(2)\hookrightarrow \cdots
 \end{equation}
where $\mathfrak{g}=\mathfrak{u}, \mathfrak{so}~ \text{or}~\mathfrak{sp}$. So, an element $Z\in \mathfrak{g}(N)$ can always be upgraded to $Z\in \mathfrak{g}(M),~M>N$ by placing $Z$ in the upper left of the matrix and filling the rest, up to dimension $M$, with 0's. Related to this embedding, there is a natural set of operators:
\begin{equation}\label{proj}
\text{Proj}_{NM}: \mathfrak{g}(M) \to  \mathfrak{g}(N).
\end{equation}
Operators $\text{Proj}_{NM}$  reduce the dimension of the matrices from $M$ to $N$ by killing the `extra' zeros of the
embedding. We can complete the definition of $\text{Proj}_{NM}$ by sending to 0 all elements $Z\in\mathfrak{g}(M)$ such that the number of eigenvalues of $Z$ which are different from 0 is greater than $N$.\\
We may easily extend the definition of $\text{Proj}_{NM}$ to act on gauge invariant operators. We will make  $\text{Proj}_{NM}$ to act as above on every $Z$ of the composite operators. This way, $\text{Proj}_{NM}$ is a map that takes gauge invariant operators built on $Z\in \mathfrak{g}(M)$ to gauge invariant operators built on $Z\in \mathfrak{g}(N)$.\\
 In terms of symmetric polynomials in the eigenvalues of $Z$, (\ref{proj}) turns into
\begin{equation*}
\text{Proj}_{NM}:~ p_{\sigma}(\lambda_1,\dots,\lambda_N,{\bf 0}_{M-N})\to p_{\sigma}(\lambda_1,\dots,\lambda_N).
\end{equation*}
At this point it is useful to think of gauge invariant operators built on $Z\in \mathfrak{g}(N)$ as vectors belonging to a vector space $V_N$. The one-to-one correspondence between states and operators in conformal field theories makes this association quite natural. Then $\text{Proj}_{NM}:V_M\to V_N$,  and the free field two-point function is an inner product defined in each $V_N$.  
 It is natural to wonder about the adjoint operators of $\text{Proj}_{NM}$ with respect to this inner product. We will call them $\text{Av}_{MN}$, and they map  gauge invariant operators  built on $Z\in \mathfrak{g}(N)$ into gauge invariant operators built on $Z\in \mathfrak{g}(M)$. So, the two point function must fulfill
\begin{equation}\label{compatibility}
\langle\text{Av}_{MN}[\mathcal{O}_N]\mathcal{O'}^+_M\rangle=\langle\mathcal{O}_N \text{Proj}_{NM}\mathcal{O'}^+_M\rangle,
\end{equation}
for $\mathcal{O}$ and $\mathcal{O'}$ arbitrary gauge invariant operators.\\
The averaging operator can be constructed as
\begin{equation}\label{Av}
\text{Av}_{MN}\mathcal{O}_N=\int_{g\in G(M)} \text{d}g~ \text{Ad}_g (\mathcal{O}_N),
\end{equation}
where $G=U, SO ~\text{or}~Sp$. $\text{d}g$ is the Haar measure of the corresponding group and $\text{Ad}_g$ is the adjoint action of $g$. The adjoint action on $Z\in \mathfrak{g}(N)\hookrightarrow\mathfrak{g}(M)$ is defined as usual:
\begin{equation*}
\text{Ad}_g(Z)=gZg^{-1},\quad g\in G(M).
\end{equation*}
Note that when $\text{Ad}_g$ is applied to a trace it does not modify the range of the indeces which were originally traced, for instance
\begin{equation}\label{exad}
\text{Ad}_g\text{Tr}_N(Z)=\text{Tr}_N(gZg^{-1})\neq \text{Tr}_N(Z),
\end{equation}
unless $M=N$, in which case, the adjoint action on gauge invariant operators is trivial.\\
As example (\ref{exad}) shows, given a gauge invariant operator $\mathcal{O}$, $\text{Ad}_g \mathcal{O}$ is in general not gauge invariant. However, 
the integral over the group restores gauge invariance. So, $\text{Av}_{MN}$, as defined in (\ref{Av}), is actually a map between gauge invariant operators.\\
Relation (\ref{compatibility}) with the definition of $\text{Av}_{MN}$ as in (\ref{Av}) will be proved for each gauge group in the next subsections. Actually, by linearity, it will be enough to prove (\ref{compatibility}) for arbitrary multitrace monomials.\\ 
Note that correlators in (\ref{compatibility}) are in different spaces: in the LHS operators are built on elements $Z\in \mathfrak{g}(M)$ whereas in the RHS $Z\in \mathfrak{g}(N)$. Equation (\ref{compatibility}) shows that $\text{AV}_{MN}=\text{Proj*}_{NM},~ \forall M,N$ with respect to the free field two-point function of the theory, as we claim. It also shows the compatibility between Weingarten and Wick calculus. To see this let us consider two operators built on  $Z\in \mathfrak{g}(N)$, for example two multitrace monomials $\text{Tr}_N(\sigma Z)$ and  $\text{Tr}_N(\tau Z)$. One can compute the correlator of these to operators as usual, summing all possible Wick contractions. This way we get a result on the RHS of equation (\ref{compatibility}). Alternatively, we can upgrade $Z\in \mathfrak{g}(N)$ so that $Z\in \mathfrak{g}(N)\hookrightarrow\mathfrak{g}(M)$ and go to the LHS of (\ref{compatibility}). One of the multitrace monomials keeps its structure, except for $Z\in \mathfrak{g}(M)$. However, the other multitrace monomial is affected by  $\text{AV}_{MN}$ and turns into a complicated sum of multitrace monomials built on  $Z\in \mathfrak{g}(M)$, as can be seen in (\ref{AvovertraceU}), (\ref{AvovertraceO}) and (\ref{averagingovertracesSp}). The spectrum of this sum comes from the integrals involved in the definition of $\text{AV}_{MN}$, that is, from Weingarten calculus. For different $M$ we get a different sum. But as relation (\ref{compatibility}) states, all these sums must be arranged in a way so that they keep the same value for the two-point function. In this sense we say that Weingarten and Wick calculus are compatible.

Now, we will consider the composition
\begin{equation}\label{composition}
\text{Proj}_{NM} \circ\text{Av}_{MN}.
\end{equation}
By construction, operators (\ref{composition}) are self-adjoint with respect to our correlators and they map gauge invariant operators built on $Z\in \mathfrak{g}(N)$ into gauge invariant operators built on $Z\in \mathfrak{g}(N)$ .\\
It is logical to wonder about the eigenvectors and the eigenvalues of (\ref{composition}). It turns out that the eigenvectors are precisely Schurs. Specifically, we will prove for each gauge group in the following subsections that\footnote{Within the context of symmetric functions, equation (\ref{eigenvectors}) first appeared in \cite{OO1,OO2} under the name of `coherence property' . Their purpose was to give a characterization of Schur polynomials. Although their definitions for $\text{Proj}_{NM}$ and $\text{Av}_{MN}$ are different from ours we have decided to keep their notation.}
\begin{equation}\label{eigenvectors}
(\text{Proj}_{NM} \circ\text{Av}_{MN})\chi_{R,N}=\frac{f_R^{G(N)}}{f_R^{G(M)}}\chi_{R,N}, \quad \forall M>N,
\end{equation}
for $G=U, SO~\text{and}~Sp$.\\
Since the eigenvalues in (\ref{eigenvectors}) are all different for different $\chi$'s and because $\text{Proj}_{NM} \circ\text{Av}_{MN}$ are self-adjoint, we conclude that Schur operators are orthogonal for classical gauge groups of any rank.\\

Note that from (\ref{compatibility}) and (\ref{eigenvectors}) one can recover the precise form of the correlator of Schurs up to a constant. That is, we can obtain
 \begin{equation}\label{cRortho}
\langle\chi_{R,N}\chi^+_{S,N}\rangle=c_Rf_R^{G(N)}\delta_{RS}.
\end{equation}
The method is quite simple. From (\ref{eigenvectors}) we know that 
\begin{equation}\label{averaging}
\text{Av}_{MN}\chi_{R,N}=\frac{f_R^{G(N)}}{f_R^{G(M)}}\chi_{R,M}, \quad \forall M>N.
\end{equation}
Applying (\ref{compatibility})
to Schur operators and using (\ref{averaging}) we get
\begin{equation*}
\frac{1}{f_R^{G(M)}}\langle\chi_{R,M}\chi^+_{R,M}\rangle=\frac{1}{f_R^{G(N)}}\langle\chi_{R,N}\chi^+_{R,N}\rangle.
\end{equation*}
This means that 
\begin{equation}\label{cR}
c_R\equiv \frac{1}{f_R^{G(M)}}\langle\chi_{R,M}\chi^+_{R,M}\rangle
\end{equation}
is finite and does not depend on $M$. Therefore, $c_R$ is a number.
Now, the orthogonality relation (\ref{cRortho}) follows from the orthogonality of Schurs and (\ref{cR}).
We may conclude that from (\ref{compatibility}) and (\ref{eigenvectors}) we recover the two point function up
to a constant. But this was expected. It is clear that both (\ref{compatibility}) and (\ref{eigenvectors}) still hold for
$c_R\chi_R$, so the freedom of multiplying every Schur operator by a constant  should be reflected in the  two
point function. Relation (\ref{cRortho}) precisely reflects this arbitrariness.\\
In order to find $c_R=1~ \forall R$ one must invariably get some result from Wick contractions.  Equation (\ref{cR}) shows that $\langle\chi_{R,M}\chi^+_{R,M}\rangle$ is proportional to $f_R^{G(M)}$. Now,  by definition, we know that polynomials $f_R^{G(M)}$ all have coefficient 1 in the highest power of $M$. So, the value of $c_R$ is the coefficient of the highest power of $M$ in the polynomial $\langle\chi_{R,M}\chi^+_{R,M}\rangle$. Therefore, the input we need from Wick contractions is, merely, that this coefficient is always equal to one.\\

Let us summarize the logic of this construction. The starting point is to extract some information of gauge invariant operators
built on $n$ fields $Z\in \mathfrak{g}(N)$. We decide to fix $n$ but move on $N$. First, we realize that the algebras can be embedded as in (\ref{embebding}). Then we think of the most basic non-trivial set of operators that adapts to this embedding and
find $\text{Proj}_{NM}$. These operators map gauge invariant operators built on $Z\in \mathfrak{g}(M)$ into gauge invariant operators built on $Z\in \mathfrak{g}(N)$. Considering gauge invariant operators as vectors and the two-point function as the inner product of the theory, we wonder which operators are the adjoints of $\text{Proj}_{NM}$, we call them $\text{Av}_{NM}$ and find that they can be constructed as in (\ref{Av}). The fact that $\text{Av}_{MN}=\text{Proj*}_{NM}$ is shown in relation (\ref{compatibility}). Moreover, we construct a set of self-adjoint operators by composition of them in (\ref{composition}) and calculate their eigenvectors and eigenvalues in (\ref{eigenvectors}). It turns out that their eigenvectors are Schur operators whose eigenvalues are all different. From there, we conclude that Schur operators form an orthogonal set.\\
To actually say that Schur operators form a basis of half-BPS gauge invariant operators we must either evaluate the partition function and match the counting, or prove that orthogonal Schurs generate all gauge invariant operators. If we assume that all gauge invariant operators are linear combinations of multitrace monomials it is easy to see that Schur operators form a basis. We should think of gauge invariant operators as symmetric polynomials in the eigenvalues of $Z$. Then the relation between symmetric polynomials
\begin{equation*}
p_\sigma(\lambda_1,\dots,\lambda_N)=\sum_R\chi_R(\sigma)s_R(\lambda_1,\dots,\lambda_N),
\end{equation*}
which is the inversion of (\ref{defSchur}), 
tells us that all multitrace monomials can be expressed in terms of Schurs, and so Schurs generate all gauge invariant operators.
This is the case of $U(N)$. However, for the gauge groups $SO(N)$ with $N$ even, we know that there is another sector of gauge invariant operators built on multitraces and pfaffians \cite{CDD}, and they cannot be generated by Schurs. So, to be precise, in those cases we must say that Schur operators form an orthogonal  basis of multitrace gauge invariant operators. \\    
The next subsections are devoted to a detailed proof of (\ref{compatibility}) and (\ref{eigenvectors}) for the different gauge groups.

\subsection{Case $U(N)$}\label{caseU}
Let us apply the general method for the case of unitary groups. Choose $N<M$ for the rank of the groups and use 
\begin{equation*}
\text{Av}_{_{MN}}(\cdot)=\int_{g\in U(M)} \text{d}g~ \text{Ad}_g (\cdot).
\end{equation*}
As explained before this operation converts gauge invariant operators built on matrices of dimension $N$ to gauge invariant operators with matrices of dimension $M$, keeping the number of fields fixed.
The integral can be performed using a result by Collins\cite{C}, which we rewrite for our convenience as:
\begin{equation}\label{IntU}
\int_{g\in U(M)} \text{d}g \ g^{i_1}_{j_1}\cdots g^{i_n}_{j_n} (g^+)^{ j'_1}_{i'_1}\cdots (g^+)^{j'_n}_{i'_n}=\sum_{\alpha, \beta \in S_n}(\alpha)^I_{I'}(\beta)^{J'}_J\text{Wg}^{U(M)}(\alpha\beta).
\end{equation}
Now, 
\begin{eqnarray}\label{Adtrace}
\text{Ad}_g (\text{Tr}_N(\sigma Z))&=&\text{Tr}_N[\sigma (gZg^{-1})]=\text{Tr}_N[\sigma (gZg^+)]\nonumber \\
&=& g^{i_1}_{j_1}Z^{j_1}_{j'_1}(g^+)^{j'_1}_{i_{\sigma(1)}}\cdots g^{i_n}_{j_n}Z^{j_n}_{j'_n}(g^+)^{j'_n}_{i_{\sigma(n)}}.
\end{eqnarray}
Remember that since $Z\in \mathfrak{u}(N)\hookrightarrow \mathfrak{u}(M)$, indeces $j,j'=1,\dots,M$, whereas $i=1,\dots,N$.
So, applying (\ref{IntU}), we find
\begin{equation}\label{AvovertraceU}
\text{Av}_{_{MN}}(\text{Tr}_N(\sigma Z))=\sum_{\alpha, \beta \in S_n}\text{Tr}_N(\sigma\alpha)\text{Tr}_M(\beta Z)\text{Wg}^{U(M)}(\alpha\beta).
\end{equation}
Now, 
\begin{equation*}
\langle\text{Av}_{MN}[\text{Tr}_N(\sigma Z)]\text{Tr}_M(\tau \bar{Z})\rangle=\sum_{\alpha, \beta \in S_n}\text{Wg}^{U(M)}(\alpha\beta)\text{Tr}_N(\sigma\alpha)\langle\text{Tr}_M(\beta Z) \text{Tr}_M(\tau \bar{Z})\rangle,
\end{equation*}
substituting the value of the Weingarten function (\ref{WgU}) and evaluating the correlator with (\ref{expandU})
\begin{equation*}
=\frac{1}{n!}\sum_{\alpha, \beta \in S_n}\sum_{\substack{\lambda \vdash n \\
l(\lambda)\leq M}}\sum_{\mu\vdash n}d_{\lambda}\frac{f_{\mu}^{U(M)}}{f_{\lambda}^{U(M)}}\text{Tr}_N(\sigma\alpha)\chi_{\lambda}(\alpha\beta)\chi_\mu(\beta)\chi_\mu(\tau),
\end{equation*}
summing over $\beta$
\begin{equation*}
=\sum_{\alpha \in S_n}\sum_{\substack{\lambda \vdash n \\
l(\lambda)\leq M}}\text{Tr}_N(\sigma\alpha)\chi_{\lambda}(\alpha)\chi_\lambda(\tau),
\end{equation*}
and this is easily seen to be
\begin{equation*}
=\langle\text{Tr}_N(\sigma Z) \text{Tr}_N(\tau \bar{Z})\rangle=\langle\text{Tr}_N(\sigma Z) \text{Proj}_{NM}[\text{Tr}_M(\tau \bar{Z})]\rangle,
\end{equation*}
since $Z\in \mathfrak{u}(N)\hookrightarrow \mathfrak{u}(M)$. This proves that $\text{Av}_{MN}=\text{Proj*}_{NM}$.\\
In the following we will focus on proving that  Schur operators are a  basis of eigenvectors of $\text{Proj}_{NM}\circ\text{Av}_{MN}$ operators for all $N$ and $M$.\\
\begin{equation}\label{AvU}
\text{Av}_{{MN}}\chi_{R,N}(Z)=\int_{g\in U(M)} \text{d}g \frac{1}{n!}\sum_{\sigma \in S_n}\chi_R(\sigma)\text{tr}_N[\sigma (gZg^{-1})],
\end{equation}
now we can use (\ref{AvovertraceU}) to write
\begin{equation*}
=\frac{1}{n!}\sum_{\alpha, \beta,\sigma \in S_n}\text{Wg}^{U(M)}(\alpha\beta)\chi_R(\sigma)\text{Tr}_N(\sigma\alpha)\text{Tr}_M(\beta Z),
\end{equation*}
after substitution of the value of $\text{Wg}^{U(M)}$ 
\begin{equation*}
=\frac{1}{(n!)^2}\sum_{\alpha, \beta,\sigma \in S_n}\sum_{\substack{\lambda \vdash n \\
l(\lambda)\leq M}}\frac{d_{\lambda}}{f_{\lambda}^{U(M)}}\chi_{\lambda}(\alpha \beta)
\chi_R(\sigma)\text{Tr}_N(\sigma\alpha)\text{Tr}_M(\beta Z),
\end{equation*}
summing over $\sigma$
\begin{equation*}
=\frac{1}{(n!)^2}\sum_{\alpha, \beta \in S_n}\sum_{\substack{\lambda \vdash n \\
l(\lambda)\leq M}}d_{\lambda}\frac{f_{\lambda}^{U(N)}}{f_{\lambda}^{U(M)}}\chi_{\lambda}(\alpha \beta)
\chi_R(\alpha)\text{Tr}_M(\beta Z),
\end{equation*}
summing over $\alpha$
\begin{equation*}
=\frac{1}{n!}\sum_{\beta \in S_n}\frac{f_{R}^{U(N)}}{f_R^{U(M)}}\chi_R(\beta)
\text{Tr}_M(\beta Z)=\frac{f_{R}^{U(N)}}{f_R^{U(M)}}\chi_{R,M}(Z).
\end{equation*}
So, 
\begin{equation*}
(\text{Proj}_{NM}\circ \text{Av}_{{MN}})\chi_{R,N}(Z)=\frac{f_{R}^{U(N)}}{f_R^{U(M)}}\chi_{R,N}(Z),
\end{equation*}
and it has been proved that Schur operators are the eigenvectors of operators $\text{Proj}_{NM}\circ \text{Av}_{{MN}}$.

\subsection{Case SO(N)}\label{caseO}
Choose $M>N$. The averaging operator is now 
\begin{equation*}
\text{Av}_{{MN}}(\cdot)=\int_{g\in O(M)} \text{d}g~ \text{Ad}_g (\cdot).
\end{equation*}
Again, this operation converts gauge invariant operators built on $Z\in \mathfrak{so}(N)$ into gauge invariant operators built on $Z\in \mathfrak{so}(M)$, keeping the number of fields fixed.
The integral can be performed using a result in \cite{CM}, which we rewrite for our convenience as:
\begin{equation}\label{IntO}
\int_{g\in O(M)} \text{d}g \ g_{i_1 j_1}\cdots g_{i_{2n} j_{2n}} =\frac{1}{(2^n n!)^2}\sum_{\alpha, \beta \in S_{2n}}\delta_{\alpha(I)}\delta_{\beta(J)}\text{Wg}^{O(M)}(\alpha^{-1}\beta),
\end{equation}
where the Weingarten function was defined in (\ref{WgO}).\\
Let us see how the adjoint action acts on traces. It is instructive to see how it works on a single trace of length 2. For convenience it will be written as
\begin{equation}\label{tr2}
\text{Ad}_g(\text{Tr}_N (Z^2))=g_{i_1 j_1}Z_{j_1 j_2}g^{-1}_{j_2 i_2}g_{i_3 j_3}Z_{j_3 j_4}g^{-1}_{j_4 i_4}\delta_{i_2 i_3}\delta_{i_4 i_1}.
\end{equation}
The indices have been chosen for the integral to apply straightforwardly. Note that because $g^{-1}=g^t$ for the elements of the orthogonal group, the indices in terms of $g$ in (\ref{tr2}) appear kindly ordered as in (\ref{IntO}).\\
 Using the shorthand notation
\begin{equation*}
(gZg)_{i_1 i_2}=g_{i_1 j_1}Z_{j_1 j_2}g_{i_2 j_2},
\end{equation*}
the adjoint action on a general multitrace monomial reads
\begin{equation*}\label{multiAdg}
\text{Ad}_g(\text{Tr}_N \sigma(Z^2)^{\otimes \frac{n}{2}}) 
=(gZg)_{I}\delta_{w_\sigma(I)}.
\end{equation*}
Now, 
\begin{equation*}
\text{Av}_{_{MN}}(\text{Tr}_N \sigma(Z^2)^{\otimes \frac{n}{2}})
=\int_{g\in O(M)} \text{d}g~ (gZg)_{I}\delta_{w_{\sigma}(I)},
\end{equation*}
by computing the integral according to (\ref{IntO})
\begin{equation*}
=\frac{1}{(2^n n!)^2}\sum_{\alpha, \beta \in S_{2n}}\text{Wg}^{O(M)}(\alpha^{-1} \beta)\delta_{\alpha(I)} \delta_{w_{\sigma}(I)} \delta_J  Z_{\beta^{-1}(J)}
\end{equation*}
\begin{equation*}
=\frac{1}{(2^n n!)^2}\sum_{\alpha, \beta \in S_{2n}}\text{Wg}^{O(M)}(\alpha^{-1} \beta)\delta_{w_{\sigma}^{-1}\alpha(I)} \delta_I  \delta_J Z_{\beta^{-1}(J)},
\end{equation*}
by expanding $\delta_{w_{\sigma}^{-1}\alpha(I)} \delta_I$ according to (\ref{traceO})
\begin{equation*}
=\frac{1}{2^n n!(2n)!}\sum_{\alpha, \beta \in S_{2n}}\sum_{\substack{\lambda \vdash n \\
l(\lambda)\leq N}}\text{Wg}^{O(M)}(\alpha^{-1} \beta)d_{2\lambda}Z_{\lambda}({\bf 1}_N)\omega_{\lambda}(w_{\sigma}^{-1}\alpha)   \delta_J Z_{\beta^{-1}(J)},
\end{equation*}
now we substitute the value of $\text{Wg}^{O(M)}$
\begin{equation*}
=\frac{1}{(2n)!^2}\sum_{\alpha, \beta \in S_{2n}}\sum_{\substack{\lambda,\mu \vdash n \\
l(\lambda)\leq N, l(\mu)\leq M}}d_{2\lambda}d_{2\mu}\frac{Z_{\lambda}({\bf 1}_N)}{Z_{\mu}({\bf 1}_{M})}\omega_{\lambda}(w_{\sigma}^{-1}\alpha)\omega_{\mu}(\alpha^{-1}\beta) \delta_J Z_{\beta^{-1}(J)}. 
\end{equation*}
Finally we sum over $\alpha$ to obtain
\begin{equation}\label{AvovertraceO}
\text{Av}_{_{MN}}(\text{Tr}_N \sigma(Z^2)^{\otimes \frac{n}{2}})=\frac{1}{(2n)!}\sum_{\beta \in S_{2n}}\sum_{\substack{\lambda \vdash n \\
l(\lambda)\leq N}}d_{2\lambda}\frac{Z_{\lambda}({\bf 1}_N)}{Z_{\lambda}({\bf 1}_{M})}\omega_{\lambda}(w_{\sigma}^{-1}\beta) \delta_J Z_{\beta^{-1}(J)}.
\end{equation}
\\
Now, we can compute
\begin{equation*}
\langle\text{Av}_{_{MN}}(\text{Tr}_N \sigma(Z^2)^{\otimes \frac{n}{2}})\text{Tr}_M (\tau(\bar{Z}^2)^{\otimes \frac{n}{2}})\rangle, 
\end{equation*}
by inserting (\ref{AvovertraceO}) and expressing multitrace monomials in our shorthand notation
\begin{equation*}
=\frac{1}{(2n)!}\sum_{\beta \in S_{2n}}\sum_{\substack{\lambda \vdash n \\
l(\lambda)\leq N}}d_{2\lambda}\frac{Z_{\lambda}({\bf 1}_N)}{Z_{\lambda}({\bf 1}_M)}\omega_{\lambda}(w_{\sigma}^{-1}\beta) \delta_{\beta(J)}\delta_{w_{\tau}(I)}  \langle Z_J \bar{Z}_I\rangle, 
\end{equation*}
evaluating the correlator
\begin{equation*}
=\frac{1}{(2n)!}\sum_{\substack{\beta \in S_{2n}\\
\xi \in S_n[S_2]}}\sum_{\substack{\lambda \vdash n \\
l(\lambda)\leq N}}d_{2\lambda}\frac{Z_{\lambda}({\bf 1}_N)}{Z_{\lambda}({\bf 1}_M)}\omega_{\lambda}(w_{\sigma}^{-1}\beta) \delta_{\beta(J)}\delta_{w_{\tau}(I)} (\xi)^J_I \text{sgn}(\xi),
\end{equation*}
and contracting indices
\begin{equation*}
=\frac{1}{(2n)!}\sum_{\substack{\beta \in S_{2n}\\
\xi \in S_n[S_2]}}\sum_{\substack{\lambda \vdash n \\
l(\lambda)\leq N}}d_{2\lambda}\frac{Z_{\lambda}({\bf 1}_N)}{Z_{\lambda}({\bf 1}_M)}\omega_{\lambda}(w_{\sigma}^{-1}\beta) \delta_{w_{\tau}^{-1}\xi \beta(I)} \delta_I \text{sgn}(\xi), 
\end{equation*}
now, we expand the trace $\delta_{w_{\tau}^{-1}\xi \beta(I)} \delta_I$ acoording to (\ref{traceO})
\begin{equation*}
=\frac{2^n n!}{(2n)!^2}\sum_{\substack{\beta \in S_{2n}\\
\xi \in S_n[S_2]}}\sum_{\substack{\lambda, \mu \vdash n \\
l(\lambda)\leq N,l(\mu)\leq M}}d_{2\lambda}d_{2\mu}\frac{Z_{\lambda}({\bf 1}_N)Z_{\mu}({\bf 1}_M)}{Z_{\lambda}({\bf 1}_M)}\omega_{\lambda}(w_{\sigma}^{-1}\beta) \omega_{\mu}(w_{\tau}^{-1}\xi \beta)\text{sgn}(\xi),
\end{equation*}
we sum over $\beta$
\begin{equation*}
=\frac{2^n n!}{(2n)!}\sum_{\xi \in S_n[S_2]}\sum_{\substack{\lambda \vdash n \\
l(\lambda)\leq N}}d_{2\lambda}Z_{\lambda}({\bf 1}_N)\omega_{\lambda}(w_{\sigma}^{-1}\xi w_{\tau}^{-1})  \text{sgn}(\xi),
\end{equation*}
We now apply the identity (\ref{Ivanovw}), and because $\lambda$ has to have even number of boxes in each column, we may say
\begin{equation*}
=\sum_{\substack{\lambda \vdash n/2\\
l(\lambda)\leq N}}2^{l(\sigma)+l(\tau)}Z_{\lambda\cup\lambda}({\bf 1}_N)\chi_{\lambda}(\sigma^{-1})\chi_{\lambda}(\tau),
\end{equation*}
or in terms of $f_{\lambda}^{SO(N)}$, as defined in (\ref{fRO})
\begin{equation*}
=\sum_{\substack{\lambda \vdash n/2\\
l(\lambda)\leq N}}2^{l(\sigma)+l(\tau)}f_{\lambda}^{SO(N)}\chi_{\lambda}(\sigma^{-1})\chi_{\lambda}(\tau) ,
\end{equation*}
but this is the correlator of multitrace monomials, so 
\begin{equation*}
=\langle\text{Tr}_N (\sigma(Z^2)^{\otimes \frac{n}{2}})\text{Tr}_N (\tau(\bar{Z}^2)^{\otimes \frac{n}{2}}\rangle,
\end{equation*}
and, finally, because $Z\in \mathfrak{so}(N)\hookrightarrow \mathfrak{so}(M)$, we may conclude
\begin{equation*}
=\langle\text{Tr}_N (\sigma(Z^2)^{\otimes \frac{n}{2}})\text{Proj}_{NM}(\text{Tr}_M (\tau(\bar{Z}^2)^{\otimes \frac{n}{2}})\rangle.
\end{equation*}
So, for orthogonal groups it is also true that $\text{Av}_{MN}=\text{Proj*}_{NM}$.
\\

Now let us prove that Schur operators are eigenvalues of $\text{Proj}_{NM}\circ\text{Av}_{MN}$ for $M>N$. We will apply first $\text{Av}_{MN}$ on $(n/2)!\chi_{R,N}(Z^2)$, so we do not have to carry a global factor all over the calculation.
\begin{equation*}
\text{Av}_{MN}[(n/2)!\chi_{R,N}(Z^2)]
=\sum_{\sigma \in S_{n/2}}2^{-l(\sigma)}\chi_R(\sigma)\text{Av}_{MN}[\text{Tr}_N (\sigma(Z^2)^{\otimes \frac{n}{2}})], \end{equation*}
applying (\ref{AvovertraceO})
\begin{equation*}
=\frac{1}{(2n)!}\sum_{\substack{\beta \in S_{2n} \\
\sigma \in S_{n/2}}}\sum_{\substack{\lambda \vdash n \\
l(\lambda)\leq N}}d_{2\lambda}\frac{Z_{\lambda}({\bf 1}_N)}{Z_{\lambda}({\bf 1}_{M})}2^{-l(\sigma)}\chi_R(\sigma)\omega_{\lambda}(w_{\sigma}^{-1}\beta) \delta_J Z_{\beta^{-1}(J)},
\end{equation*}
using the symmetry (\ref{symmetryO}) and averaging over $S_n[S_2]$
\begin{equation*}
=\frac{1}{(2n)!2^n n!}\sum_{\substack{\beta \in S_{2n} \\
\sigma \in S_{n/2}\\
\xi \in S_n[S_2]}}\sum_{\substack{\lambda \vdash n \\
l(\lambda)\leq N}}d_{2\lambda}\frac{Z_{\lambda}({\bf 1}_N)}{Z_{\lambda}({\bf 1}_{M})}2^{-l(\sigma)}\chi_R(\sigma)\omega_{\lambda}(w_{\sigma}^{-1}\beta) \delta_{\xi\beta(J)} Z_J \text{sgn}(\xi)
\end{equation*}
\begin{equation*}
=\frac{1}{(2n)!2^n n!}\sum_{\substack{\beta \in S_{2n} \\
\sigma \in S_{n/2}\\
\xi \in S_n[S_2]}}\sum_{\substack{\lambda \vdash n \\
l(\lambda)\leq N}}d_{2\lambda}\frac{Z_{\lambda}({\bf 1}_N)}{Z_{\lambda}({\bf 1}_{M})}2^{-l(\sigma)}\chi_R(\sigma)\omega_{\lambda}(w_{\sigma}^{-1}\xi\beta)\text{sgn}(\xi) \delta_{\beta(J)} Z_J,
\end{equation*}
the properties of (\ref{Ivanov2}) allow us to take representatives for $\beta$ and sum over the elements of the double coset  
\begin{equation*}
=\frac{2^n n!}{(2n)!}\sum_{\substack{\nu \vdash n/2 \\
\sigma \in S_{n/2}\\
\xi \in S_n[S_2]}}\sum_{\substack{\lambda \vdash n \\
l(\lambda)\leq N}}d_{2\lambda}\frac{Z_{\lambda}({\bf 1}_N)}{Z_{\lambda}({\bf 1}_{M})}2^{-l(\sigma)}\chi_R(\sigma)z^{-1}_{4\nu}\omega_{\lambda}(w_{\sigma}^{-1}\xi w_{\nu})\text{sgn}(\xi) \delta_{w_{\nu}(J)} Z_J ,
\end{equation*}
after using $z_{4\mu}=2^{2l(\mu)}z_{\mu}$ and writing in the notation of traces
\begin{equation*}
=\frac{2^n n!}{(2n)!}\sum_{\substack{\nu \vdash n/2 \\
\sigma \in S_{n/2}\\
\xi \in S_n[S_2]}}\sum_{\substack{\lambda \vdash n \\
l(\lambda)\leq N}}d_{2\lambda}\frac{Z_{\lambda}({\bf 1}_N)}{Z_{\lambda}({\bf 1}_{M})}2^{-l(\sigma)-2l(\nu)}\chi_R(\sigma) z^{-1}_{\nu}\omega_{\lambda}(w_{\sigma}^{-1}\xi w_{\nu})\text{sgn}(\xi)\text{Tr}_M (\nu(Z^2)^{\otimes \frac{n}{2}}),
\end{equation*}
if we apply again (\ref{Ivanov2})
\begin{equation*}
=\sum_{\substack{\nu \vdash n/2 \\
\sigma \in S_{n/2}}}\sum_{\substack{\lambda \vdash n/2 \\
l(\lambda)\leq N}}\frac{Z_{\lambda\cup\lambda}({\bf 1}_N)}{Z_{\lambda\cup\lambda}({\bf 1}_{M})}2^{-l(\nu)}\chi_R(\sigma)\chi_{\lambda}(\sigma)\chi_{\lambda}(\nu) z^{-1}_{\nu}\text{Tr}_M (\nu(Z^2)^{\otimes \frac{n}{2}}),
\end{equation*}
since all the dependence on $\mu$ is through class functions we may actually sum over elements of $\tau\in S_{n/2}$ 
\begin{equation*}
=\frac{1}{(n/2)!}\sum_{\tau,
\sigma \in S_{n/2}}\sum_{\substack{\lambda \vdash n/2 \\
l(\lambda)\leq N}}\frac{Z_{\lambda\cup\lambda}({\bf 1}_N)}{Z_{\lambda\cup\lambda}({\bf 1}_{M})}2^{-l(\tau)}\chi_R(\sigma)\chi_{\lambda}(\sigma)\chi_{\lambda}(\tau) \text{Tr}_M (\tau(Z^2)^{\otimes \frac{n}{2}}),
\end{equation*}
and summing over $\sigma$
\begin{equation*}
=\sum_{\tau, \in S_{n/2}}\frac{Z_{R\cup R}({\bf 1}_{N})}{Z_{R\cup R}({\bf 1}_{M})}2^{-l(\tau)}\chi_{R}(\tau) \text{Tr}_M (\tau(Z^2)^{\otimes \frac{n}{2}}), 
\end{equation*}
which is actually
\begin{equation*}
=\frac{f_R^{SO(N)}}{f_R^{SO(M)}}(n/2)!\chi_{R,M}(Z^2).
\end{equation*}
As before, since $Z\in \mathfrak{so}(N)\hookrightarrow \mathfrak{so}(M)$, operator $\text{Proj}_{NM}$ acts by simply changing $M\to N$ in the label of Schurs. Thus
\begin{equation}
(\text{Proj}_{NM}\circ\text{Av}_{MN})\chi_{R,N}(Z^2)=\frac{f_R^{SO(N)}}{f_R^{SO(M)}}\chi_{R,N}(Z^2),
\end{equation}
and the statement is proved.

\subsection{Case $Sp(N)$}
Choose $M>N$. The averaging operator is now 
\begin{equation*}
\text{Av}_{{MN}}(\cdot)=\int_{g\in Sp(M)} \text{d}g~ \text{Ad}_g (\cdot).
\end{equation*}
Again, this operation converts gauge invariant operators built on $Z\in \mathfrak{sp}(N)$ into gauge invariant operators built on $Z\in \mathfrak{sp}(M)$, keeping the number of fields fixed.
The integral can be performed using a result in \cite{Mat}, which we rewrite for our convenience as:
\begin{equation}\label{IntSp}
\int_{g\in Sp(M)} \text{d}g \ g_{i_1 j_1}\cdots g_{i_{2n} j_{2n}} =\frac{1}{(2^n n!)^2}\sum_{\alpha, \beta \in S_{2n}}J_{\alpha(I)}J_{\beta(J)}\text{Wg}^{Sp(M)}(\alpha^{-1}\beta),
\end{equation}
where Weingarten functions where given in (\ref{WgSp}).
\\
Using (\ref{traceJ}) and for $w=(1234)$, we have for the case of two fields:  
\begin{equation*}
\text{Ad}_g(\text{Tr}_N(Z^2))=J_I\text{Ad}_g(ZJ)_{w(I)}=J_I(gZg^{-1}J)_{w(I)}=J_{w^{-1}(I)}(gZJg^t)_{I}.
\end{equation*}
It is convenient to write
\begin{equation}\label{kordered}
(gZJg^t)_{I}=g_{i_1 j_1} (ZJ)_{j_1 j_2} g_{i_2 j_2}~ g_{i_3 j_3} (ZJ)_{j_3 j_4} g_{i_4 j_4}.
\end{equation}
In general 
\begin{equation*}
\text{Ad}_g[\text{Tr}_N(\sigma(Z^2)^{\otimes\frac{n}{2}} )]=J_{w_\sigma^{-1}(I)}(gZJg^t)_{I},
\end{equation*}
and $(gZJg^t)_{I}$ can be written as in (\ref{kordered}).
\\
Now the averaging operator acting on traces can be computed:
\begin{equation*}
\text{Av}_{MN}[\text{Tr}_N(\sigma(Z^2)^{\otimes\frac{n}{2}} )]=\int_{g\in Sp(M)} \text{d}g~ J_{w_\sigma^{-1}(I)}(gZJg^t)_{I},
\end{equation*}
and implementing (\ref{IntSp}) 
\begin{equation*}
=\frac{1}{(2^n n!)^2}\sum_{\alpha, \beta \in S_{2n}}\text{Wg}^{Sp(M)}(\alpha^{-1}\beta)J_{w_\sigma^{-1}(I)}J_{\alpha(I)}J_{\beta(J)}(ZJ)_J
\end{equation*}
\begin{equation*}
=\frac{1}{(2^n n!)^2}\sum_{\alpha, \beta \in S_{2n}}\text{Wg}^{Sp(M)}(\alpha^{-1}\beta)J_IJ_{\alpha w_\sigma(I)}J_{\beta(J)}(ZJ)_J,
\end{equation*}
if we expand the trace $J_IJ_{\alpha w_\sigma(I)}$ according to (\ref{traceSp})
\begin{equation*}
=\frac{1}{2^n n!(2n)!}\sum_{\substack{\lambda \vdash n \\
l(\lambda)\leq N}}\sum_{\alpha, \beta \in S_{2n}}\text{Wg}^{Sp(M)}(\alpha^{-1}\beta)d_{\lambda\cup\lambda}Z'_{\lambda}({\bf 1}_{N/2})\omega^\varepsilon_{\lambda}(\alpha w_\sigma)J_{\beta(J)}(ZJ)_J,
\end{equation*}
now we substitute the value of $\text{Wg}^{Sp(M)}$ according to (\ref{WgSp})
\begin{equation*}
=\frac{1}{(2n)!^2}\sum_{\substack{\lambda,\mu \vdash n \\
l(\lambda)\leq N}}\sum_{\alpha, \beta \in S_{2n}}d_{\lambda\cup\lambda}d_{\mu\cup\mu}\frac{Z'_{\lambda}({\bf 1}_{N/2})}{Z'_{\mu}({\bf 1}_{M/2})}\omega^\varepsilon_{\lambda}(\alpha w_\sigma)\omega^\varepsilon_{\mu}(\alpha^{-1}\beta)J_{\beta(J)}(ZJ)_J,
\end{equation*}
and if we sum over $\alpha$ 
\begin{equation*}
=\frac{1}{(2n)!}\sum_{\substack{\lambda \vdash n \\
l(\lambda)\leq N}}\sum_{\beta \in S_{2n}}d_{\lambda\cup\lambda}\frac{Z'_{\lambda}({\bf 1}_{N/2})}{Z'_{\lambda}({\bf 1}_{M/2})}\omega^\varepsilon_{\lambda}(w_\sigma \beta)J_{\beta(J)}(ZJ)_J,
\end{equation*}
since $J_{\beta(J)}(ZJ)_J=J_{\eta\beta(J)}(ZJ)_J$, with $\eta\in S_n[S_2]$, we can average over $\eta$
\begin{equation*}
=\frac{1}{(2n)!2^n n!}\sum_{\substack{\lambda \vdash n \\
l(\lambda)\leq N}}\sum_{\substack{\beta \in S_{2n}\\
\eta\in S_n[S_2]}}d_{\lambda\cup\lambda}\frac{Z'_{\lambda}({\bf 1}_{N/2})}{Z'_{\lambda}({\bf 1}_{M/2})}\omega^\varepsilon_{\lambda}(w_\sigma \beta)J_{\eta\beta(J)}(ZJ)_J
\end{equation*}
\begin{equation*}
=\frac{1}{(2n)!2^n n!}\sum_{\substack{\lambda \vdash n \\
l(\lambda)\leq N}}\sum_{\substack{\beta \in S_{2n}\\
\eta\in S_n[S_2]}}d_{\lambda\cup\lambda}\frac{Z'_{\lambda}({\bf 1}_{N/2})}{Z'_{\lambda}({\bf 1}_{M/2})}\omega^\varepsilon_{\lambda}(w_\sigma \eta^{-1}\beta)J_{\beta(J)}(ZJ)_J,
\end{equation*}
taking representatives of $\beta$ and summing over the double coset
\begin{equation*}
=\frac{(2^n n!)}{(2n)!}\sum_{\substack{\lambda \vdash n \\
l(\lambda)\leq N}}\sum_{\substack{\mu\vdash n/2\\
\eta\in S_n[S_2]}}d_{\lambda\cup\lambda}\frac{Z'_{\lambda}({\bf 1}_{N/2})}{Z'_{\lambda}({\bf 1}_{M/2})}z_{4\mu}^{-1}\omega^\varepsilon_{\lambda}(w_\sigma \eta^{-1}w_\mu)J_{w_\mu(J)}(ZJ)_J,
\end{equation*}
applying (\ref{Ivanovw}) for twisted spherical functions
\begin{equation*}
=\sum_{\substack{\lambda \vdash n/2 \\
l(\lambda)\leq N}}\sum_{\mu\vdash n/2}\frac{Z'_{2\lambda}({\bf 1}_{N/2})}{Z'_{2\lambda}({\bf 1}_{M/2})}z_{4\mu}^{-1}2^{l(\sigma)+l(\mu)}\chi_\lambda(\sigma) \chi_\lambda(\mu)J_{w_\mu(J)}(ZJ)_J,
\end{equation*}
so, in terms of traces and elements of $S_{n/2}$ we have
\begin{eqnarray}\label{averagingovertracesSp}
&&\text{Av}_{MN}[\text{Tr}_N(\sigma(Z^2)^{\otimes\frac{n}{2}} )] \nonumber \\
&&=2^{l(\sigma)}\frac{1}{(n/2)!}\sum_{\substack{\lambda \vdash n/2\\
\mu\in S_{ n/2} \\
l(\lambda)\leq N}}\frac{Z'_{2\lambda}({\bf 1}_{N/2})}{Z'_{2\lambda}({\bf 1}_{M/2})} \chi_\lambda(\sigma)2^{-l(\mu)}\chi_\lambda(\mu)\text{Tr}_M(\mu(Z^2)^{\otimes\frac{n}{2}} ).
\end{eqnarray}
Now, we can compute
\begin{equation*}
\langle\text{Av}_{{MN}}(\text{Tr}_N \sigma(Z^2)^{\otimes \frac{n}{2}})\text{Tr}_M (\tau(\bar{Z}^2)^{\otimes \frac{n}{2}})\rangle, 
\end{equation*}
by inserting (\ref{averagingovertracesSp}) 
\begin{equation*}
=\frac{1}{(n/2)!}\sum_{\substack{\lambda \vdash n/2\\
\mu\in S_{ n/2} \\
l(\lambda)\leq N}}\frac{Z'_{2\lambda}({\bf 1}_{N/2})}{Z'_{2\lambda}({\bf 1}_{M/2})}2^{l(\sigma)-l(\mu)}\chi_\lambda(\sigma) \chi_\lambda(\mu)\langle\text{Tr}_M(\mu(Z^2)^{\otimes\frac{n}{2}})\text{Tr}_M (\tau(\bar{Z}^2)^{\otimes \frac{n}{2}})\rangle,
\end{equation*}
making use of (\ref{coortraceSp})
\begin{equation*}
=\frac{1}{(n/2)!}\sum_{\substack{\lambda,\nu \vdash n/2\\
\mu\in S_{ n/2} \\
l(\lambda)\leq N}}\frac{Z'_{2\lambda}({\bf 1}_{N/2})}{Z'_{2\lambda}({\bf 1}_{M/2})}2^{l(\tau)+l(\sigma)}\chi_\lambda(\sigma) \chi_\lambda(\mu)\chi_\nu(\mu)\chi_\nu(\tau)f_\nu^{Sp(M)},
\end{equation*}
summing over $\mu$ and applying (\ref{fRsPZ})
\begin{equation*}
=2^{l(\sigma)+l(\tau)}\sum_{\substack{\lambda \vdash n/2 \\
l(\lambda)\leq N}}\chi_\lambda(\sigma)\chi_\lambda(\tau)f_\lambda^{Sp(N)},
\end{equation*}
but this is precisely
\begin{equation*}
=\langle\text{Tr}_N (\sigma(Z^2)^{\otimes \frac{n}{2}})\text{Proj}_{NM}[\text{Tr}_M (\tau(\bar{Z}^2)^{\otimes \frac{n}{2}})]\rangle. 
\end{equation*}
Thus $\text{Av}_{MN}=\text{Proj*}_{NM}$ with respect to our correlators in the symplectic case as well. 
\\

To prove that Schur operators built on elements of $\mathfrak{sp}(N)$ are also eigenvectors of $\text{Proj}_{NM}\circ\text{Av}_{MN}$ we use (\ref{averagingovertracesSp}) and write
\begin{equation*}
 \text{Av}_{MN}[\chi_{R,N}(Z^2)]=\frac{1}{(n/2)!^2}\sum_{\substack{\lambda \vdash n/2\\
\mu,\sigma\in S_{ n/2} \\
l(\lambda)\leq N}}\frac{Z'_{2\lambda}({\bf 1}_{N/2})}{Z'_{2\lambda}({\bf 1}_{M/2})}\chi_R(\sigma) \chi_\lambda(\sigma)2^{-l(\mu)}\chi_\lambda(\mu)\text{Tr}_M(\mu(Z^2)^{\otimes\frac{n}{2}} ),
\end{equation*}
which, after summing over $\sigma$ is
\begin{equation*}
=\frac{1}{(n/2)!}\sum_{
\mu\in S_{ n/2} }\frac{Z'_{2R}({\bf 1}_{N/2})}{Z'_{2R}({\bf 1}_{M/2})}2^{-l(\mu)}\chi_R(\mu)\text{Tr}_M(\mu(Z^2)^{\otimes\frac{n}{2}} ),
\end{equation*}
and after applying (\ref{fRsPZ}) 
\begin{equation*}
=\frac{f_R^{Sp(N)}}{f_R^{Sp(M)}}\frac{1}{(n/2)!}\sum_{
\mu\in S_{ n/2}}2^{-l(\mu)}\chi_R(\mu)\text{Tr}_M(\mu(Z^2)^{\otimes\frac{n}{2}} )=\frac{f_R^{Sp(N)}}{f_R^{Sp(M)}}\chi_{R,M}(Z^2).
\end{equation*}
As before, operators $\text{Proj}_{NM}$  act naturally on operators built on $Z\in \mathfrak{sp}(N)\hookrightarrow \mathfrak{sp}(M)$, so
\begin{equation*}
(\text{Proj}_{NM}\circ\text{Av}_{MN})\chi_{R,N}(Z^2)=\frac{f_R^{Sp(N)}}{f_R^{Sp(M)}}\chi_{R,N}(Z^2),
\end{equation*}
and orthogonality follows immediately since the eigenvalues are all different.

\section{Conclusions and future works}\label{Con}
In this paper we have shown how an orthogonal basis of Schurs for half-BPS gauge invariant operators emerges naturally from the embedding structure of the algebras. This point of view allows us to treat all classical gauge groups in a unified framework and gives a neat answer to the question of why Schur operators are orthogonal for all classical gauge groups, as found in \cite{CDD2}. The main results of the paper are equations (\ref{compatibility}) and (\ref{eigenvectors}).\\
Operators $\text{Av}$ and $\text{Proj}$, which we use to translate the structure of embedding algebras into maps between half-BPS gauge invariant operators, can naturally act on gauge invariant operators built on more that one matrix. One of the immediate extensions of this work is to verify that (\ref{compatibility}) and (\ref{eigenvectors}) still hold for restricted Schurs and explore the consequences. In other words, to answer the question of what the embedding algebra structure can tell about restricted Schurs.\\
The techniques we need to state and prove our results are, as far as we know, new in this field. The use of Weingarten calculus, for instance, makes contact with random matrix theory. It would be interesting to explore some connections in that line. \\
In our opinion, it is very suggesting that important results can come out from a preexisting structure. This method could be extrapolate to other structures. For example, if instead of fixing the number of fields and move on the rank parameter we do the other way round, the structure would no longer be of embedding algebras but of a Bratteli diagram of the symmetric group. This should contain some information about the dilatation operator.

\section*{Acknowledgements}
The author would like to thank Robert de Mello Koch whose ideas and suggestions have helped to enrich this paper. The author is also grateful to Hesam Soltanpanahi Sarabi and Pawel Caputa for their useful comments after reading the manuscript. This work has been partly supported by a Claude Leon Fellowship.

\end{document}